\documentclass[preprint,sort&compress]{elsarticle}
\usepackage{amsmath}
\usepackage{amsfonts}
\usepackage{amssymb}
\usepackage{epsfig}
\usepackage{natbib}
\usepackage{comment}

\usepackage{numcompress}

\journal{Computer Physics Communications}

\begin{document}

\begin{frontmatter}

\title{GPU accelerated Trotter-Suzuki solver for quantum spin dynamics}

\author[ifeg,famaf,invap]{Axel~D.~Dente}
\ead{dente@famaf.unc.edu.ar}

\author[ifeg,famaf]{Carlos~S.~Bederi\'{a}n}
\ead{bc@famaf.unc.edu.ar}

\author[ifeg,famaf]{Pablo~R.~Zangara}

\author[ifeg,famaf]{Horacio~M.~Pastawski}
\ead{horacio@famaf.unc.edu.ar}

\address[ifeg]{Instituto de F\'{i}sica Enrique Gaviola (IFEG), CONICET-UNC, 5000, C\'{o}rdoba, Argentina}
\address[famaf]{Facultad de Matem\'{a}tica, Astronom\'{i}a y F\'{i}sica, Universidad Nacional de C\'{o}rdoba, 5000, C\'{o}rdoba, Argentina}
\address[invap]{INVAP S.E., 8403, San Carlos de Bariloche, Argentina}

\begin{abstract}

The resolution of dynamics in out of equilibrium quantum spin systems lies at the heart of fundamental questions among Quantum Information Processing, Statistical Mechanics and Nano-Technologies. Efficient computational simulations of interacting many-spin systems are extremely valuable tools for tackling such questions. Here, we use the Trotter-Suzuki (TS) algorithm, a well-known strategy that provides the evolution of quantum systems, to address the spin dynamics. We present a GPU implementation of a particular TS version, which has been previously implemented on single cores in CPUs. We develop a massive parallel version of this algorithm and compare the efficiency between CPU and GPU implementations. This method reduces the execution time considerably and is capable of dealing with systems of up to 27 spins (only limited by GPU memory).

\end{abstract}

\end{frontmatter}


\section{Introduction\label{Sec: Introduction}}

The evolution of a generic quantum spin system is described by an appropriate
Schr\"{o}dinger equation, where the Hamiltonian operator $H$ encloses all
the information about its dynamics, including external fields and spin-spin
interactions. The ``academic'' strategy involves the solution of an eigen-problem
for $H$ and the state vectors $\left\vert \psi\right\rangle $
\cite{merzbacher-1998-quantum-Mechanics}. This can be a major computational task, since the dimensionality
of the underlying Hilbert space scales exponentially with the number $N$ of interacting spins in the system.
In practice, such scenario constrains the problem to the treatment of
few-spins systems, or dealing with special cases where symmetries provide
further simplifications \cite{LiebMattis-1961}.

Nowadays, there are many alternative strategies that can be employed instead of the full matrix diagonalization. A reliable approximate method needs to be unitary, i.e. it must conserve the total probability during the quantum evolution. In particular, standard numerical integrators like the Runge-Kutta algorithm fail to fulfill such condition. By contrast, the Trotter-Suzuki (TS) algorithm \cite{HdeRaedt-TrotterApplications,DeRaedt-trotter} does preserve unitarity, since it approximates the exact evolution operator by a set of adequately built partial evolution operators. Provided that the time steps are short enough, the exact evolution is well approximated by the successive partial evolutions. Additionally, it is worthy to mention that the TS method does not rely on any Hilbert space truncation. This is particularly important when addressing long-time asymptotics of interacting many-body systems, a problem of major relevance in fundamental physics \cite{rigolNATURE2008,polkovnikovRMP}.

Within the last years, Graphics Processing Units (GPU) have been successfully employed to accelerate numerical algorithms that solve the Schr\"{o}dinger equation  \cite{Bederian-Dente-Trotter,Cucchietti,Bauke}. Most of these implementations deal with one-body or wave-packet quantum dynamics. However, the potential of the GPU has not been exploited within a wide range of many-body problems, such as interacting spin systems. Here, spin-spin interactions provide a substantial complexity, which turns out to be a crucial challenge not only for the computational implementation but also for out-of-equilibrium physics.

In this article we report on an implementation of a 4th order TS decomposition in GPU. We consider
intrinsically interacting spin Hamiltonians without any Hilbert space truncation or any further
assumption about symmetries or ad hoc dynamical conditions. Our implementation is based on the XYZ decomposition \cite{DeRaedt-XYZ-Descomposition}, in which the Hamiltonian is partitioned in the X, Y and Z components of the bilinear spin interactions. Since the Z component can be written in a simple diagonal form, it only modifies the phases of the states. Moreover, since the X and Y components are non-diagonal, local rotations are performed in order to map them into Z-like terms.

The Z-like terms and the local rotations are implemented by means of a massive parallelization scheme. The speedup obtained goes beyond 30 times compared with an OpenMP implementation on a 6-core CPU. Despite this decrease in execution time, the system size is still bounded by the maximum amount of memory available, which is about 6GB for current-generation GPUs. Thus, our implementation is capable of reaching a system of 27 spins, which is still a significant number when compared to other methods.

The paper is organized as follows. In Sec. \ref{Sec:Trotter algorithm} we summarize
the basis of the TS algorithm. In Sec. \ref{Sec:Implementation} we describe the main procedures performed by our GPU implementation. In Sec. \ref{Sec:Results} we discuss the performance and accuracy.  We include also in \ref{Sec:Entangled-State} a brief discussion on two scenarios where this computational strategy can be employed: magnetic resonance spin ensemble calculations, where the present work is currently being exploited, and its potential use in the evolution of Matrix Product States. Sec. \ref{Sec:Conclusions} concludes the present work.

\section{The Trotter-Suzuki Algorithm \label{Sec:Trotter algorithm}}

In this section we summarize the TS method for spin systems as
presented in Ref. \cite{DeRaedt-trotter}. We consider a time-independent spin $\frac{1}{2}$ Hamiltonian:

\begin{equation}
H={\displaystyle\sum\limits_{j=1}^{N}}
{\displaystyle\sum\limits_{\alpha=x,y,z}}
h_{j}^{\alpha}\text{\ }S_{j}^{\alpha}+%
{\displaystyle\sum\limits_{j,k=1}^{N}}
{\displaystyle\sum\limits_{\alpha=x,y,z}}
J_{j,k}^{\alpha}S_{j}^{\alpha}S_{k}^{\alpha}\text{,}%
\label{Eq: Hamiltoniano General}%
\end{equation}
where $S_{j}^{\alpha}$ is the spin operator at site $j$ and $\alpha=x,y,z$. The
parameters $h_{j}^{\alpha}$ define the local fields or chemical shifts, with the constraint $\langle h_{j}^{\alpha} \rangle \equiv 0$,  while $J_{j,k}^{\alpha}$ are the couplings between two spins.

We assume that the $ N $-spin system is initially described by a state vector $\left\vert
\Psi_{0}\right\rangle $, which is expanded in the computational Ising basis:

\begin{equation}
\left\vert
\Psi_{0}\right\rangle = {\displaystyle\sum\limits_{i=1}^{2^{N}}} c_{i}\left\vert \beta_{i}\right\rangle \text{.}
\label{Eq: inicial}%
\end{equation}
Here, $c_{i}$ are complex coefficients and $ \left\vert \beta_{i}\right\rangle $ are tensor products of eigenvectors of each $S_{j}^{z}$, typically referred as the $S^{z}$-\textit{decoupled} or \textit{Ising} basis.  In \ref{Sec:Entangled-State} we discuss two particularly relevant cases in which \textit{pure} states, as in Eq. \ref{Eq: inicial}, are employed to evaluate the dynamics of complex many-body systems. In fact, we stress
that the present computational strategy can be substantially exploited when combined with sophisticated physically-based
representations \cite{Alvarez-PRL-2008}.

The state of the system at an arbitrary time $t$ is formally given by  $\left\vert \Psi_{t}\right\rangle = \mathbf{U}(t)\left\vert \Psi_{0}\right\rangle = e^{-\mathrm{i}tH/\hbar} \left\vert \Psi_{0}\right\rangle$. For simplicity, we set from now on $\hbar=1$.

The main idea of the TS method relies on finding an appropriate partition of $H$ which may yield a set of simple partial evolutions that approximate the exact one $\mathbf{U}(t)$. If we consider Eq. \ref{Eq: Hamiltoniano General} as a particular sum of terms $H= H_{1}+...+H_{K} $, then:

\[
\mathbf{U}(t)=e^{-\mathrm{i}tH}=e^{-\mathrm{i}t\left(  H_{1}+...+H_{K}\right)
}=\lim_{m\rightarrow\infty}\left(  \prod_{k}^{K}
e^{-\frac{\mathrm{i}tH_{k}}{m}}\right)  ^{m} \text{.}
\]

The first order approximation to $\mathbf{U}(t)$ is%

\begin{equation}
\mathbf{U}(t) \backsimeq \mathbf{\tilde{U}}_{\mathbf{1}}(t)=e^{-\mathrm{i}%
tH_{1}}...e^{-\mathrm{i}tH_{K}}\text{,}%
\label{Eq: U1}
\end{equation}
while the second and fourth order approximations can be expressed in terms of the first:
\[
\mathbf{\tilde{U}}_{\mathbf{2}}(t)=\mathbf{\tilde{U}}_{\mathbf{1}}^{\dag
}(-t/2)\mathbf{\tilde{U}}_{\mathbf{1}}(t/2)\text{,}%
\]
\[
\mathbf{\tilde{U}}_{\mathbf{4}}(t)=\mathbf{\tilde{U}}_{\mathbf{2}%
}(at)\mathbf{\tilde{U}}_{\mathbf{2}}(at)\mathbf{\tilde{U}}_{\mathbf{2}%
}((1-4a)t)\mathbf{\tilde{U}}_{\mathbf{2}}(at)\mathbf{\tilde{U}}_{\mathbf{2}%
}(at)\text{,}%
\]
where $a=1/(4-4^{1/3})$.
These are indeed satisfactory approximations to $\mathbf{U}(t)$ provided that $t$ must be small enough compared to the maximum local time-scale determined by $H$. Such role is played by the local second moment of the interactions. This means that the partial evolution time step $ t $ must satisfy:

\begin{equation}
t \ll \left[ \underset{j,k}{max}\Vert H_{j,k} \Vert \right] ^{-1} \simeq \left[  \underset{j,k}{max} \left[ {\displaystyle\sum\limits_{\alpha=x,y,z}}
\left(\frac{h_{j}^{\alpha}}{2}\right)^2 +%
{\displaystyle\sum\limits_{\alpha=x,y,z}}
 \left( \frac{J_{j,k}^{\alpha}}{4} \right)^2 \right]^{\frac{1}{2}} \right]^{-1}
\label{Eq: SegundoMomento}
\end{equation}

Now we address specifically how to build an operator $\mathbf{\tilde{U}}_{\mathbf{1}}(t)$ given by Eq. \ref{Eq: U1} from the total Hamiltonian of Eq. \ref{Eq: Hamiltoniano General}. The natural choice for the partition sum are the single-spin and the two-spin terms, which shall be mapped into a diagonal representation by suitable rotations. In particular, we consider
the operators that rotate $X$ and $Y$ axis into the $Z$ axis:

\begin{equation}
R^{y}_{\pi/2} = \frac{1}{\sqrt{2}}
\begin{bmatrix} 1 & -1 \\ 1 & 1 \end{bmatrix}, \\
R^{x}_{-\pi/2} = \frac{1}{\sqrt{2}}
\begin{bmatrix} 1 & i \\ i & 1 \end{bmatrix}, \label{Eq:Rotations}
\end{equation}

and satisfy:

\[
(R^{y}_{\pi/2})^{\dagger}S^{x}R^{y}_{\pi/2} = S^{z},
\]

\[
(R^{x}_{-\pi/2})^{\dagger}S^{y}R^{x}_{-\pi/2}  = S^{z}.
\]

These operators rotate $S^{x}$ and $S^{y}$ into $S^{z}$ with the purpose of performing any phase correction in the computational Ising basis. The partial evolution $ \exp\left[ -\mathrm{i}t h_{j}^{\alpha}S_{j}^{\alpha}\right] $ yields a trivial phase for $\alpha = z$, while $\alpha = x,y$ need to be properly rotated using $ R^{y}_{\pi/2,j} $ and  $ R^{x}_{-\pi/2,j} $ respectively. The $ j$-index labels each spin, and the corresponding global rotations are defined by $Y=\bigotimes_{j}R^{y}_{\pi/2,j}$ and $X=\bigotimes_{j}R^{x}_{-\pi/2,j}$.

Let us first consider the single-spin operations,

\begin{equation}
\exp\left(  -\mathrm{i}t\left[
{\displaystyle\sum\limits_{j=1}^{N}}{\displaystyle\sum\limits_{\alpha =x,y,z}}h_{j}^{\alpha}
S_{j}^{\alpha}\right]  \right)  \backsimeq
\prod_{\alpha=x,y,z}\exp\left[
-\mathrm{i}t{\displaystyle\sum\limits_{j =1}^{N}}h_{j}^{\alpha}S_{j}^{\alpha}\right]. 
\label{Eq: SPIN Rotation}%
\end{equation}
Here, we stress that the approximate equality relies on the validity of Eq. \ref{Eq: SegundoMomento}. As mentioned above, non trivial exponentials are rotated:

\begin{align}
\exp\left[-\mathrm{i}t \displaystyle\sum\limits_{j=1}^{N} h_{j}^{x} S_{j}^{x}\right]   &  = Y \exp\left(  -\mathrm{i} t \displaystyle\sum\limits_{j=1}^{N} h_{j}^{x}S_{j}^{z}  \right)  Y^{\dagger}, %
\nonumber\\
\exp\left[-\mathrm{i}t \displaystyle\sum\limits_{j=1}^{N} h_{j}^{y} S_{j}^{y}\right]      &  =X \exp\left(  -\mathrm{i} t \displaystyle\sum\limits_{j=1}^{N} h_{j}^{y}S_{j}^{z}  \right)  X^{\dagger}.%
\label{Eq: Single Spin Rotation}
\end{align}

Notice that this kind of single-spin operations can be computed exactly without rotations and without the approximation of Eq. \ref{Eq: SPIN Rotation}. Nevertheless, our purpose is to write all the partial evolution operators in an explicit diagonal form. This strategy
will be specifically exploited in the computational implementation, as shown in Section \ref{Sec:Implementation}.

In analogy with Eq. \ref{Eq: SPIN Rotation}, the two-spin operators yield:

\begin{align}
\exp\left(  -\mathrm{i}t\left[
{\displaystyle\sum\limits_{j,k=1}^{N}}
{\displaystyle\sum\limits_{\alpha=x,y,z}}
J_{j,k}^{\alpha}S_{j}^{\alpha}S_{k}^{\alpha}\right]  \right)  \simeq  \prod_{\alpha=x,y,z} \exp\left(  -\mathrm{i}t\left[
{\displaystyle\sum\limits_{j,k=1}^{N}}
J_{j,k}^{\alpha}S_{j}^{\alpha}S_{k}^{\alpha}\right]  \right)
\end{align}

Once again, the Z-terms yield diagonal operators:

\begin{equation}
e^{it J_{j,k}^z S_j^z S_k^z} =
\begin{bmatrix} e^{it J_{j,k}^z /4} & & & \\ & e^{-it J_{j,k}^z /4} & &  \\ & &e^{-it J_{j,k}^z /4} & \\ & & & e^{it J_{j,k}^z /4}\end{bmatrix}.
\label{Eq:ZZ-Phases}
\end{equation}

The remaining two-spin terms are rotated by $Y$ and $X$ accordingly,

\begin{align}
\exp\left(  -\mathrm{i}t\left[{\displaystyle\sum\limits_{j,k=1}^{N}}
J_{j,k}^{\alpha}S_{j}^{x}S_{k}^{x}\right]  \right) = %
Y \exp\left(  -\mathrm{i}t\left[{\displaystyle\sum\limits_{j,k=1}^{N}}
J_{j,k}^{x}S_{j}^{z}S_{k}^{z}\right] \right) Y^{\dagger},%
\nonumber\\
\exp\left(  -\mathrm{i}t\left[{\displaystyle\sum\limits_{j,k=1}^{N}}
J_{j,k}^{\alpha}S_{j}^{y}S_{k}^{y}\right]  \right) = %
X \exp\left(  -\mathrm{i}t\left[{\displaystyle\sum\limits_{j,k=1}^{N}}
J_{j,k}^{x}S_{j}^{z}S_{k}^{z}\right] \right) X^{\dagger}.%
\label{Eq: Two-Spin-Operators}
\end{align}

Notice that, from Eqs. \ref{Eq: Single Spin Rotation} and \ref{Eq: Two-Spin-Operators}, the rotations for single and double spin terms can be performed simultaneously. In such sense, we stress that the TS implementation described here has been intentionally prepared in order to enable parallelization.

\section{Implementation \label{Sec:Implementation}}

Any state written in the form of Eq. \ref{Eq: inicial} can be evolved by the successive application of
partial evolution operators, as those defined in Sec. \ref{Sec:Trotter algorithm}. At any time, the state vector is represented by a
double precision complex array that occupies $2^{N}\times16$ bytes of memory. Its $k$th element stores the probability amplitude $ c_{k} $ for the corresponding state of the Ising basis. Since these $2^{N}$ states are tensor products of up and down configurations for each spin, they can be written according to the $N$-bit binary representation of $k$. Naturally, the size of the system is constrained by the maximum amount of memory available in the GPU (in our case, 6GB for a NVIDIA Tesla C2075).

The implementation of the evolution is divided into two main modules:
phase-corrections and axis-rotations.%

\[
\text{Forward Rotation }\rightarrow\text{Phases in }Z\rightarrow\text{
Backward Rotation}%
\]

\subsection{Phase correction}
This module performs the phase corrections depending on the Z-projection (up or down) and position of each spin in each
state of the Ising basis. In this stage there are no cross terms between different basis states (all operations are diagonal), and therefore the parallelization is trivial. However, due to the two-spin terms, the time of evaluation of each phase correction increases quadratically with the number of spins.

Since the Hamiltonian is time-independent, the phase corrections can be computed in advance (i.e. before performing the actual sequence of partial evolutions) and stored in memory for the following evolutions, amortizing its high cost. In fact, Eq. \ref{Eq:ZZ-Phases} ensures that the phases only depend on the time step and the Hamiltonian parameters. The disadvantage of such pre-computed phase corrections lies on the amount of memory required to store three $2^{N}$-element vectors, increasing memory usage by 150\% and limiting simulations to 27 spins on 6GB GPUs instead of 28. This technique, however, reduces phase correction time for large systems by 97\%.

Additionally, since every phase correction is followed by a backward rotation that also operates on the whole state,
the phase correction kernel is merged with the backward rotation kernel. Therefore, the phase correction is performed when
the rotation kernel reads elements from memory and just before applying the rotation. This saves two global memory
operations on each element and a kernel call.

Combined, these improvements reduce the overall simulation time by up to 52\%.

\subsection{Rotation}

The rotation module acts on pairs of basis states by means of the operators $X$ and $Y$ defined in Eq. \ref{Eq:Rotations}. If a particular basis state has the $j$-th spin down, then it is paired to the basis state that has the $j$-th spin up and the same configuration for the rest of spins. Let $n_{j,\downarrow}$ be the binary representation of a particular basis state in which the $j$-th spin\footnote{Consistently with the binary representation, spins are enumerated from right to left.} is down, i.e. its $j$-th bit is $0$: $ n_{j,\downarrow}=\ldots\downarrow_j\ldots $. Then, $n_{j,\downarrow}$ is paired to $n_{j,\uparrow} =\ldots\uparrow_j\ldots = n_{j,\downarrow}+2^{j}$.


This pairing procedure must range the whole Hilbert space. Thus, for every value of $j$ the pairing must be repeated over the $2^{N-1}$ states that have the sought configuration in the $j$-th spin.

In order to show how the parallelization can be performed at this stage, let us exemplify one case of the pairing procedure. If we address the rotation of the second spin in a four-spin system, then the pairs of states under consideration are the following:%

\[%
\begin{array}
[c]{c}%
\downarrow\downarrow \boldsymbol{\downarrow} \downarrow\\
\downarrow\downarrow \boldsymbol{\uparrow} \downarrow
\end{array}
,%
\begin{array}
[c]{c}%
\downarrow\downarrow \boldsymbol{\downarrow} \uparrow\\
\downarrow\downarrow \boldsymbol{\uparrow} \uparrow
\end{array}
,%
\begin{array}
[c]{c}%
\downarrow\uparrow \boldsymbol{\downarrow} \downarrow\\
\downarrow\uparrow \boldsymbol{\uparrow} \downarrow
\end{array}
,%
\begin{array}
[c]{c}%
\downarrow\uparrow \boldsymbol{\downarrow} \uparrow\\
\downarrow\uparrow \boldsymbol{\uparrow} \uparrow
\end{array}
,%
\begin{array}
[c]{c}%
\uparrow\downarrow \boldsymbol{\downarrow} \downarrow\\
\uparrow\downarrow \boldsymbol{\uparrow} \downarrow
\end{array}
,%
\begin{array}
[c]{c}%
\uparrow\downarrow \boldsymbol{\downarrow} \uparrow\\
\uparrow\downarrow \boldsymbol{\uparrow} \uparrow
\end{array}
,%
\begin{array}
[c]{c}%
\uparrow\uparrow \boldsymbol{\downarrow} \downarrow\\
\uparrow\uparrow \boldsymbol{\uparrow} \downarrow
\end{array}
\text{ and }%
\begin{array}
[c]{c}%
\uparrow\uparrow \boldsymbol{\downarrow} \uparrow\\
\uparrow\uparrow \boldsymbol{\uparrow} \uparrow
\end{array}
\text{.}%
\]

Notice that the number of pairings is $8=2^{N-1}$. From this example it is clear that there are no repetitions in any of the basis states when executing one specific rotation (i.e. a particular $j$). Thus, each rotation operation within the same $j$-spin results independent and can be parallelized.

\subsection{Efficient rotation on GPUs}

As stated above, the rotation of the $j$-th spin involves $2^{N-1}$ independent operations. Accordingly, a GPU kernel that rotates the $j$-th spin on a single pair is launched on $2^{N-1}$ threads, one per pair. Since the rotation has an extremely low arithmetic intensity of 0.125 operations per byte transferred (i.e. 8 operations per pair of states read and written back to memory) compared to our GPU's 27.46 double precision FLOPS/bandwidth ratio, the kernel is severely limited by memory bandwidth to feed each thread with data. To avoid this bottleneck, the optimized implementation reads each of these states from global memory and performs as many rotations as possible before writing them back.

Let us consider a subset of $M$ specific spins ($M\leq N$) denoted by $F$ and fix its state configuration, i.e. with a particular choice of $\uparrow$ or $ \downarrow $ for each of the $M$ spins. Then, the set of all states that satisfy the specified configuration for the spins in $F$ is closed under the pairing operation that flips any spin which is not in $F$. For example, in the case $N=4$ one can consider the $F$ set composed by spins 1 and 3 (i.e. $M=2$). Now, if we fix the configuration $\cdot \uparrow_{3} \cdot \downarrow_{1}$, the set of states that satisfy such configuration is $\lbrace 0100_{b}; 0110_{b}; 1100_{b}; 1110_{b}   \rbrace$, and this set is closed under the pairing operation for spins 2 and 4 (those which are not in $F$).

The previous observation allows us to launch $2^{M}$ thread blocks on the GPU, and define their specific spin configurations (the state of the $M$ spins in $F$) using the binary encoding of each thread block's unique index. This leaves each thread block with its own set of $2^{N - M}$ states to rotate and, by the observation above, every state's peers for spins $\{1..N\} - F$ are also within the set. Each thread block can then copy its set of states to the fast shared memory available in each multiprocessor within a GPU, and perform the rotations for all the spins in $\{1..N\} - F$ without any further global memory access. Once these rotations are done, the results are copied back to global memory and the kernel finishes.


\subsection{Coalescing memory accesses}

While the previous procedure improves rotation time, the approach has a problem: when rotating ``higher'' spins, the lower bits of the state handled by a thread block are defined by its thread block index, which results in adjacent states being interleaved across different thread blocks. For example, when rotating spins 1 and 2 in a single kernel call on a 4-spin problem, the set of states rotated by each thread block is:

\begin{center}
	\begin{tabular}{ | c | l l l l | }
		\hline
		Thread block & \multicolumn{4}{|c|}{States} \\
		\hline
		0 & $0000_b (0)$,  & $0001_b (1)$,  & $0010_b (2)$,  & $0011_b (3) $ \\
		1 & $0100_b (4)$,  & $0101_b (5)$,  & $0110_b (6)$,  & $0111_b (7) $ \\
		2 & $1000_b (8)$,  & $1001_b (9)$,  & $1010_b (10)$, & $1011_b (11) $ \\
		3 & $1100_b (12)$, & $1101_b (13)$, & $1110_b (14)$, & $1111_b (15) $ \\
		\hline
	\end{tabular}
\end{center}

However, when rotating spins 3 and 4, the set of states rotated by each thread block is:

\begin{center}
  	\begin{tabular}{ | c | l l l l |}
		\hline
		Thread block & \multicolumn{4}{|c|}{States} \\
		\hline
		0 & $0000_b (0)$, & $0100_b (4)$, & $1000_b (8)$,  & $1100_b (12) $ \\
		1 & $0001_b (1)$, & $0101_b (5)$, & $1001_b (9)$,  & $1101_b (13) $ \\
		2 & $0010_b (2)$, & $0110_b (6)$, & $1010_b (10)$, & $1110_b (14) $ \\
		3 & $0011_b (3)$, & $0111_b (7)$, & $1011_b (11)$, & $1111_b (15) $ \\
		\hline
	\end{tabular}
\end{center}

It is easy to notice that consecutive states (e.g.  $0000_b$ and $0001_b$) are assigned to different thread blocks.
Since GPU memory controllers fetch long bursts (currently 128 bytes long) of contiguous data to L1 cache, which is local to each multiprocessor (which execute few thread blocks at a time), this interleaving wastes memory bandwidth significantly.

\begin{figure}
    \centering
     \includegraphics[width=0.9\textwidth]{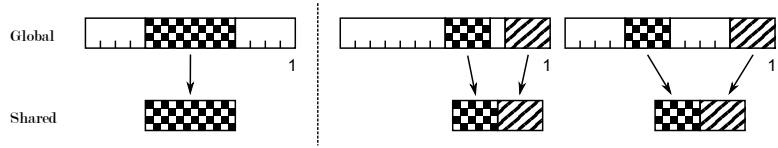}\\
      \caption{Scheme of memory accesses for a 6-spin rotation. The spins are arranged from left to right, enumerated from 1. The clear spins are set by the thread block index, and chequered spins  $\lbrace 5,6,...,10 \rbrace$ are assigned and rotated within each block.  Left panel: rotation of spins $\lbrace 5,6,...,10 \rbrace$ without memory coalescing. Right panel: the striped $s=3$ spins $\lbrace 1,2,3 \rbrace$ are used for memory coalescing; two kernel calls are performed, rotating spins $\lbrace 5,6,7 \rbrace$ and then $\lbrace 8,9,10 \rbrace$.}
    \label{Fig: Rotations}%
\end{figure}

In order to mitigate the mentioned problem, we never include the $s$ lowest spins $[1..s]$ in $F$, so that each thread block always operates on runs of at least $2^{s}$ consecutive elements in memory. Only the first rotation kernel that is called in a full evolution rotates spins $[1..s]$, while the rest of the kernels perform $s$ fewer rotations than in the previous approach because these lower spins still consume shared memory. The improved bandwidth efficiency allowed by coalesced memory accesses results in a  $300\%$ speedup over the previous kernel on high-spin rotations for $s = 3$, which matches the memory controller's 128-byte line length. Figure \ref{Fig: Rotations} shows how a 6-spin rotation is performed using one kernel call with $s = 0$ and two kernel calls with $s = 3$.

We also tried storing these $2^{s}$ elements in registers instead of shared memory, to preserve the maximum amount of rotations per kernel call. This generated significant register pressure due to the size of complex double precision numbers, producing low multiprocessor occupancy, and resulted in the kernel performing worse than the kernel that uses shared memory to store every element.

We measured the rotation times for different values of $s$ and different thread block sizes, and found out that there are no optimal values that work for every rotation and every system size. Thus, we wrote a small tool that does a brute force search for the fastest series of rotations that perform a full evolution for every system size that fits in memory. These optimal configurations are stored in a file and are used by our rotation code in following executions.

\subsection{Porting the model back to the CPU}
We replicated the GPU scheme on a CPU using OpenMP intrinsics with few changes to the algorithm: we still precompute phase corrections, divide the system into blocks and perform as many rotations as possible, but each thread loops over whole blocks at a time. The rotation code is instantiated for many different block sizes and values of $s$ using C++ template metaprogramming, providing the compiler with as much compile-time data as possible to perform automatic vectorization. Instead of using shared memory to store intermediate values, since the blocking strategy has strong temporal and spatial locality we assume that the caches on the CPU will hold the values without hitting the memory bus. As in the GPU code, optimal rotation sizes and their tunable parameters are precomputed using a separate brute force search tool.

\section{Results \label{Sec:Results}}

\subsection{Performance}

\begin{figure}
    \centering
     \includegraphics[width=0.9\textwidth]{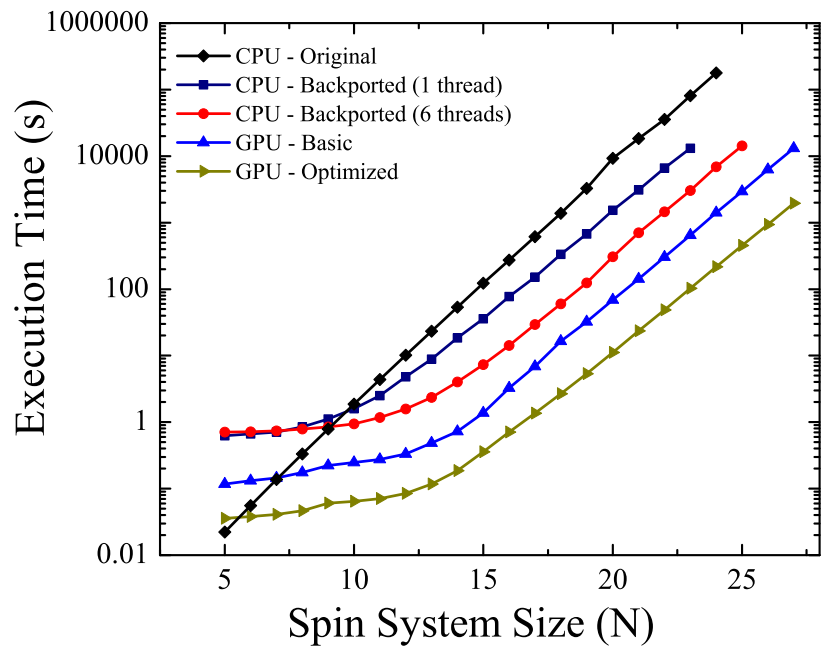}\\
      \caption{Execution time (in logarithmic scale) for CPU and GPU implementations of the Trotter-Suzuki method as a function of the number of spins. Each point correspond to the evolution with 50 Trotter-Suzuki steps.}%
    \label{Fig: CPU vs GPU}%
\end{figure}

As a result of this implementation we plot in Fig. \ref{Fig: CPU vs GPU} the execution times for the original Fortran CPU code without optimizations, the backported CPU code running on one and six threads, and the three different GPU versions. The tests were run on an Intel Core i7-980 six-core processor with triple channel DDR3-1066 memory and a Tesla C2075 GPU. The CPU code was compiled with GCC 4.7.2, using \texttt{-Ofast -march=native -mtune=native -fno-exceptions -fno-rtti -flto -fopenmp} compiler flags, while the GPU code was compiled with CUDA 5.0 using \texttt{-O -use\_fast\_math -gencode arch=compute\_20,code=sm\_20} compiler flags.

In Fig. \ref{Fig: CPU vs GPU}, it can also be observed that for small systems, the original CPU code is faster than parallel implementations because computing the quadratic phase correction is faster than performing a table lookup for small values of $N$, and it does not suffer from parallelization overhead like its CPU siblings. Meanwhile, the GPU versions do not have enough work to feed all the GPU's execution resources.

\begin{figure}
    \centering
      \includegraphics[width=0.9\textwidth]{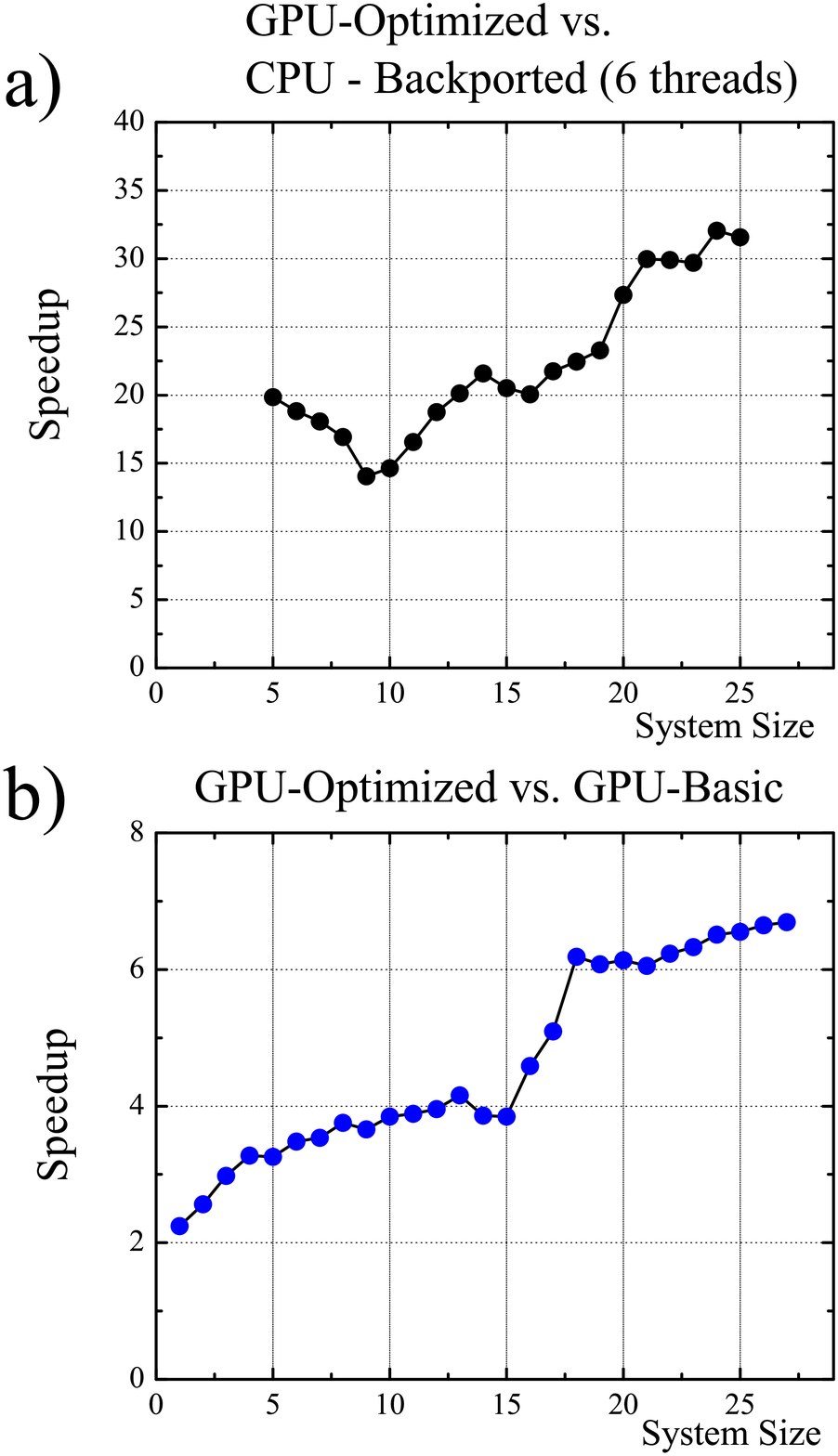}\\
      \caption{Speedup between the implementations. \textbf{a}: CPU-Backported vs GPU-Optimized. For large spin systems the GPU is about 30 times faster than the CPU. \textbf{b}: Optimized GPU version versus a naive implementation in the GPU. An increase greater than 6 times for long spin systems is observed.}%
    \label{Fig: Speedup}%
\end{figure}

On the other hand, when $N$ is considerably large, the execution time of the GPU implementation increases exponentially in $N$. The inflection point around $N\simeq14$ indicates that the GPU reaches its full capacity, and afterwards each GPU version scales as $\alpha^N$, with a factor $\alpha$ higher than $2$. In fact, this parameter can be obtained by a linear fitting of each curve in Fig. \ref{Fig: CPU vs GPU} for $N>14$. The original CPU implementation yields $\alpha=2.26$, while the CPU Backported (6 thread) yields $\alpha=2.13$. The naive GPU-Basic and the GPU-Optimized result $\alpha=2.13$ and $\alpha=2.05$ respectively. These small differences between the implementations are reflected in the behavior of the relative speedup, which is shown in Fig. \ref{Fig: Speedup}-\textbf{a} and \textbf{b} for selected implementations.

Since the memory bandwidth is the main limiting factor of the algorithm's performance, we measured the bandwidth achieved by the multiple rotation kernel to verify the quality of our implementation. Our fastest CUDA kernel implementation uses 80\% of the 120GBps reported by NVIDIA's bandwidth test tool included in the CUDA SDK, while the backported CPU implementation uses between 50\% and 70\% of the 10 GBps reported by the STREAM benchmark \cite{McCalpin2007} on our system. These results reflect our decision to focus our optimizations on GPU code and ensure that our CPU implementation avoids hitting the memory bus during a series of rotations. We estimate that better tuned CPU implementations can extract a similar percentage of the platform's memory bandwidth. However, with current high-end dual-socket quad-channel DDR3 platforms reaching 100 GBps \cite{XeonBW} and current high-end GPUs doubling this quantity \cite{TeslaBW} we expect GPUs to maintain their dominance. The high bandwidth utilization on the GPU also implies that further improvements will require a different algorithm to perform rotations.

\subsection{Accuracy}

The last issue to be addressed concerns the accuracy of our TS implementation. Several comparisons were performed between the present method and exact diagonalization schemes \cite{Lea2,Lea3,Lea4}, i.e. the solution of the Schr\"{o}dinger equation as an eigenproblem. The evolution of local and non-local physical magnitudes was compared for one-dimensional spin systems described by Hamiltonians in the form of Eq. \ref{Eq: Hamiltoniano General}. For $N=16$ \footnote{Standard exact diagonalization is limited by the size of systems that can be handled.}, a sequence of $5\times10^6$ TS steps yields a relative difference bounded by $10^{-8}$. A second accuracy test was provided by the spurious total magnetization observed in polarization-conserving Hamiltonians. In fact, this a strictly physical condition: if the initial state has zero total magnetization, then the evolved one should remain so. In general, we observe a linear increase of the total magnetization of the spin system, which is consistent with the general expectancies of a linear increase of the TS error as function of the number of steps. In particular, for $N=22$ and $5\times10^{6}$ TS steps, we observe a relative deviation less than $10^{-6}$, a precision which is good enough for practical purposes.

An alternative strategy for error quantification is provided by the \textit{Loschmidt echo} (LE) \cite{scholarpedia}. It evaluates
the reversibility of a system's dynamics in the presence of uncontrolled degrees of freedom \cite{Zangara-PRA-2012}. For a particular initial state $ \left\vert \Psi_{0}^{{}}\right\rangle $, the standard LE definition \cite{jalpa} is:

\begin{equation}
M_{LE}(2t)=\vert \left\langle \Psi_{0}^{{}} \right\vert exp\left[ \mathrm{i}(H_{0}+\Sigma)t\right]  exp\left[ -\mathrm{i}H_{0}t\right]   \left\vert \Psi_{0}^{{}}\right\rangle \vert^{2},
\end{equation}
where $ H_{0} $ is a reversed Hamiltonian, and $ \Sigma $ encloses uncontrolled non-reversed degrees of freedom. In the present case, we set $ \Sigma \equiv 0 $, which for an ideal perfectly accurate computation would yield $ M_{LE}\equiv 1 $, $ \forall t $. In principle, this statement is completely irrespective of the TS approximation and its intrinsic accuracy (i.e. the validity of Eq. \ref{Eq: SegundoMomento}). As a matter of fact, exactly the same sequences of TS partial evolutions are applied in the forward and backwards dynamics, except for the change in the sign of $ t $. Therefore, the deviations of $ M_{LE}$ away from the unity could be intrinsically originated in the execution of floating point operations by the specific Hardware.

We consider initial states given by random superpositions of Ising states (e.g. Eq. \ref{Eq: entanglado}), for a $ N=16 $ spin set. Two cases of different dynamical complexity are evaluated. The first case is a ring configuration described by a nearest neighbors Heisenberg Hamiltonian. As above, this interaction conserves total magnetization, and then the dynamics does not explore the whole Hilbert space. For this case, $5\times10^{5}$ TS steps yield $ \vert 1 - M_{LE} \vert = 5.206647 \times10^{-8} $, while $5\times10^{6}$ TS steps yield $ \vert 1 - M_{LE} \vert = 5.2066504 \times10^{-7} $. Again, the error appears to increase linearly with the number of TS steps. However, we stress here that these deviations cannot be associated to the TS decomposition. The second case is built from the first, adding double quantum terms $ S_{j}^{x} S_{k}^{x}- S_{j}^{y} S_{k}^{y}$ up to third next nearest neighbors. In this situation, total magnetization is not conserved, and therefore dynamics effectively mixes all spin projection subspaces exploring the whole Hilbert space. It turns out that $5\times10^{6}$ TS steps yield $ \vert 1 - M_{LE} \vert = 5.2094531  \times10^{-7} $, which is slightly larger than the first case. This may indicate that errors in computational operations can depend on the complexity of the many-body dynamics.

Once again, these examples evidence a satisfactory accuracy of our TS-GPU implementation. A systematic study of the LE as an error quantifier is well beyond the scope of this article. This would require ranging over $ N $ and the number of TS steps, different Hamiltonian complexities and initial states, among many other factors. However, the LE turns to be a promising witness to address computational errors in GPU and CPU implementations.

\section{Conclusions \label{Sec:Conclusions}}

We presented here a GPU implementation that boosted the Trotter-Suzuki method for quantum spin dynamics. We developed a parallelization scheme to exploit the massive parallel architecture of the GPU cards. The results showed a significant increase of performance when compared to a similar CPU implementation. In our tested platform, the speedup was measured to be of up to 30 times.

The benefits provided by this massive parallel hardware, boosted the capability of evaluating the dynamics of considerably large quantum spin systems. In particular, we were able to evolve a maximum of $27$ spins (limited only by the GPU memory) in reasonable execution times.

The comparison between our Trotter-Suzuki implementation with exact numerical approaches yielded estimated relative errors which turned to be fairly acceptable within the standard expectancies of this kind of computational strategy.

The implementation of this algorithm and the efficiency achieved open promising opportunities for studying fundamental questions within the field of out-of-equilibrium quantum many-body systems.

\section{Acknowledgements}

The authors acknowledge Lea F. Santos for providing the data for accuracy comparison, and Nicol\'{a}s Wolovick for fruitful discussions. This work was performed with the financial support from CONICET, ANPCyT, SeCyT-UNC, MinCyT-Cor, and the NVIDIA professor partnership program.

\appendix
\section{Spin systems described by single pure states \label{Sec:Entangled-State}}

We will briefly mention here two cases where an interacting many-spin system $\mathcal{S}$ can be described
or approximately described by, a pure state: the random superposition (entangled) states for high temperature systems, and the so-called Matrix Product States. These are suited candidates for our GPU-boosted TS implementation.

In the first case, $\mathcal{S}$ is characterized by the infinite temperature limit, as it is often the situation in Magnetic Resonance spin dynamics. Strictly speaking, $\mathcal{S}$ cannot be described by a pure
(single vector) state as in Eq. \ref{Eq: inicial}, but by a highly mixed state, typically denoted by a
density matrix. This represents a whole probabilistic ensemble, and contains all
the statistical information about $\mathcal{S}$ \cite{blum-1996}. The manipulation of the density
matrix may rapidly be cumbersome due to memory requirements, as soon as its
dimension scales as $2^{N}\times2^{N}$. But, provided the observables to be evaluated are local
(they involve just a few spins within $\mathcal{S}$), one can use just a few pure entangled
states to compute ensemble-averaged quantities \cite{Alvarez-PRL-2008}. This procedure enables a
physical parallelization which relies on the quantum superposition principle. A simple case may be an initial state given by a single spin up-polarized (localized excitation), and the rest of them in the high-temperature thermal
equilibrium, represented by a mixture of all states with amplitudes
satisfying the appropriate statistical weights and random
phases:

\begin{equation}\label{Eq: entanglado}
\left\vert \Psi_{0}^{{}}\right\rangle =\left\vert \uparrow\right\rangle
_{1}\otimes\left\{
{\displaystyle\sum\limits_{j=1}^{2^{N-1}}}
\frac{1}{\sqrt{2^{N-1}}}e^{\mathrm{i}\varphi_{j}}\text{\ }\left\vert
\beta_{j}\right\rangle \right\}  ,\text{ \ \ \ }\varphi_{j}^{{}}%
=\text{random phase,}%
\end{equation}

where, analogously to Eq. \ref{Eq: inicial}, $\left\vert \beta_{j}\right\rangle $ are the states of the computational Ising basis. This case has been employed to evaluate specific time-dependent correlation functions for spin systems \cite{alvarez-PRA-2010,Zangara-PRA-2012,Fine2012}, avoiding the storage, manipulation and diagonalization of overwhelmingly large matrices. Most importantly, it is nowadays employed to address the problem of thermalization in closed quantum systems, being assisted with our GPU implementation.

The second case, refers to Matrix Product States \cite{Werner1992,Banuls2009}, which constitute a set of states that successfully approximates the exact state of $\mathcal{S}$ in many physical situations. They are intimately related to renormalization group methods\cite{White-DMRG-1992}, and have proved very useful to deal with dynamical observables in one-dimensional quantum spin systems\cite{Banuls2012}.

For a chain of $ N $ spins, a Matrix Product State is given by:

\begin{equation}\label{Eq: entanglado}
\left\vert \Psi_{0}^{{}}\right\rangle =
{\displaystyle\sum\limits_{i_{1},...,i_{N}=\uparrow, \downarrow}}
tr(A_{1}^{i_{1}}...A_{N}^{i_{N}})\text{\ }\left\vert
i_{1},...,i_{N}\right\rangle %
\end{equation}

where $ A_{k}^{i} $ is a $ D $-dimensional matrix, and $ \left\vert i_{1},...,i_{N}\right\rangle $ represents an Ising state. The time-evolution of this kind of states is performed by a suitable TS algorithm\cite{Banuls2012}, and therefore they are promising candidates for TS-GPU implementations like the one we present in this article.

\bibliographystyle{elsarticle-num}

\end{document}